\documentclass[pra,a4paper,nofootinbib,twocolumn,showpacs,preprintnumbers,amsmath,amssymb,floatfix,amstex,superscriptaddress]{revtex4}
\newcommand{\ket}[1]{|#1\rangle}                        %
\newcommand{\bra}[1]{\langle #1}                        %
\usepackage{graphicx,graphics,wrapfig,rotating}         
\usepackage{dcolumn}                                    
\usepackage{bm,fancybox}                                
\usepackage{times,euscript,eufrak,oldgerm}              %
\usepackage[english]{babel}                      %
\usepackage{psfrag}
\usepackage{ulem}
\usepackage{color}

\begin{document}

\title{High-fidelity ion-trap quantum computing with hyperfine clock states}
\author{L. Aolita}
\affiliation{%
Instituto de F\'\i sica, Universidade Federal do Rio de Janeiro, Caixa Postal
68528, 21941--972 Rio de Janeiro, RJ, Brazil\\
}
\affiliation{%
Max-Planck-Institut f\"ur Physik Komplexer Systeme, N\"othnitzerstrasse 38, D-01187, Dresden, Germany\\
}
\author{K. Kim}
\affiliation{%
Institut f\"ur
Experimentalphysik, Universit\"{a}t Innsbruck, Technikerstra{\ss}e 25, A--6020
Innsbruck, Austria\\
}%
\author{J. Benhelm}
\affiliation{%
Institut f\"ur Experimentalphysik, Universit\"{a}t Innsbruck,
Technikerstra{\ss}e 25, A--6020
Innsbruck, Austria\\
}%
\author{C. F. Roos}
\affiliation{%
Institut f\"ur Experimentalphysik, Universit\"{a}t Innsbruck,
Technikerstra{\ss}e 25, A--6020
Innsbruck, Austria\\
}%
\affiliation{%
Institut f\"ur Quantenoptik und Quanteninformation der \"Osterreichischen Akademie der
Wissenschaften, Technikerstra{\ss}e 21a, A--6020 Innsbruck, Austria\\
}%
\author{H. H\"{a}ffner}
\affiliation{%
Institut f\"ur Experimentalphysik, Universit\"{a}t Innsbruck,
Technikerstra{\ss}e 25, A--6020
Innsbruck, Austria\\
}%
\affiliation{%
Institut f\"ur Quantenoptik und Quanteninformation der \"Osterreichischen Akademie der
Wissenschaften, Technikerstra{\ss}e 21a, A--6020 Innsbruck, Austria\\
}%
\date{\today}%
\definecolor{colorChristian}{rgb}{0.5,0.1,0.1}
\newcommand{\newCR}[1]{\textcolor{colorChristian}{#1}}
\newcommand{\oldCR}[1]{\textcolor{colorChristian}{\sout{#1}}}
\newcommand{\new}[1]{\textcolor{red}{#1}}
\newcommand{\context}[1]{\textcolor{magenta}{\bf #1}}
\newcommand{\newnew}[1]{\textcolor{green}{#1}}
\newcommand{\old}[1]{\textcolor{blue}{\sout{#1}}}
\newcommand{\Leandro}[1]{\textcolor{green}{\it{#1}}}
\begin{abstract}
We propose  the implementation
of a geometric-phase gate on magnetic-field-insensitive qubits
with $\hat{\sigma}^z$-dependent forces  for trapped ion quantum
computing. The force is exerted by two laser beams in a Raman configuration. Qubit-state
 dependency is achieved by a small
frequency detuning from the virtually-excited state.
Ion species with excited states of long radiative
lifetimes are used to reduce the chance of a spontaneous photon
emission to less
than 10$^{-8}$ per gate-run. This eliminates
the main source of gate infidelity of previous implementations.
With this scheme it seems possible to reach the fault
tolerant threshold.
\end{abstract}
\pacs{03.67.Lx, 03.67.Pp, 32.80.Qk}     
 \maketitle

\par
{\it Introduction}. One of the big challenges in quantum computing today is to perform
nearly perfect gate operations. With trapped ions, initialization,
read-out and single qubit operations have been already
demonstrated by various groups with very high fidelity. The
implementation of a two-qubit gate operation is more involved as
it demands  coupling the internal states of two ions
separated by a distance many orders of magnitude larger than the
Bohr radius. Ideally, the gate should operate at high speed and
with a fidelity allowing for fault-tolerant quantum computation.
To comply with the latter requirement, the gate operation should
not couple the qubits dissipatively to the environment (for
example, by spontaneous emission), nor should the qubits be
affected by dephasing mechanisms like fluctuating magnetic fields
or path lengths in the optical setup. With trapped ions impressive progress
towards high-fidelity two-qubit gates has been made over the last
couple of years, yet all current experimental realizations
\cite{Leibfried1, Riebe06,Haljan1,Home06} 
achieve fidelities not exceeding 97\% and thus still fall short of coming
close to the desired level of precision \cite{Steane}.

\par Optical Raman fields are a convenient tool to create strong
spin-dependent forces on hyperfine ion-qubits \cite{Monroe,
Leibfried1, Haljan, Haljan1}, where ``spin"  refers to the
effective Pauli spin associated with the qubit's two level system.
Acting on a single ion, these forces can entangle the
internal-spin and motional degrees of freedom of the ion
\cite{Monroe, Haljan}; following disentanglement, the ion acquires
a net geometric phase that is spin-dependent \cite{Milburn}. When
the force is exerted simultaneously on two ions, 
the geometric phase depends nonlinearly on their spins and
can be used to entangle  the ions' internal
degrees of freedom \cite{Leibfried1, Haljan1, Milburn,
Molmer-sorensen, Wang}. Quantum gates based on these collective
laser-ion interactions come in two flavours. In the first approach,
experimentally pursued by Leibfried {\it et. al.} \cite{Leibfried1},
a moving standing wave pattern of two laser beams induces, for properly
chosen laser beam polarizations, a spatially-varying state-dependent AC-Stark shift. This generates in turn a
$\sigma^z$-dependent force, {\it i. e.}, a force on the ions whose amplitude depends on the internal energy states.  The
resulting controlled-phase gate, called $\hat{\sigma}^z$-gate, is remarkably fast and robust. 

\par
\begin{figure}
\begin{center}
\includegraphics[width=1\linewidth]{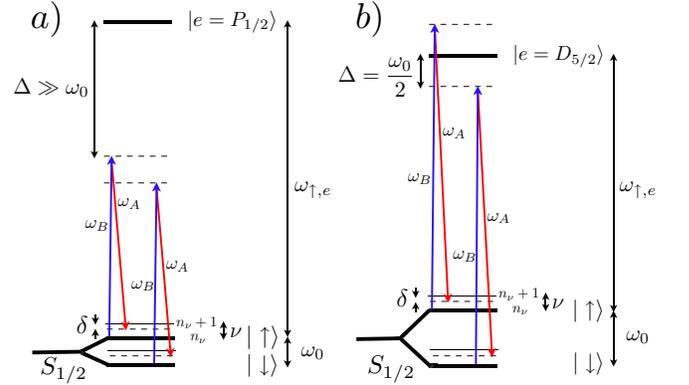}
\caption{(Not to scale, color online). The usual implementation $a)$ uses a
$P_{1/2}$ level as the excited state and works with
$\Delta\gg\omega_0$; the qubit states are magnetic-field-sensitive
and the differential coupling is achieved mainly through
polarization discrimination. In the new implementation $b)$ the
excited state is a $D_{5/2}$ metastable level and $\Delta$ is
small enough for the qubit states to be spectrally discriminated,
which allows for the use of clock states. Maximal discrimination
occurs at $\Delta=\frac{\omega_{0}}{2}$, where the effective
coupling of $\ket{\!\uparrow}$ to its motional sideband is exactly
opposite to that of $\ket{\!\downarrow}$. 
\label{frequencies}
}
\end{center}
\end{figure}

 As the Raman
coupling in \cite{Leibfried1} is mediated by a short-lived $P_{1/2}$ level (see Fig.~\ref{frequencies} $a)$), spontaneous 
emission dominates the error budget even for a large detuning $\Delta$ from the virtual
level \cite{Ozeri}.
Unfortunately,  the use of a large Raman detunning makes the gate operation very inefficient
when acting on ``clock'' states (states whose
energy splitting is first-order independent of changes of the magnetic field)
\cite{Haljan, Blinov, Lee}, for no differential Stark shift can be
induced on them in the limit 
$\Delta\gg\omega_{0}$ \cite{Lee} , where $\omega_0$ is the
hyperfine structure splitting of the qubit states. Therefore, the remarkably long coherence times \cite{Langer} available for clock states
cannot directly be combined with the intrinsic robustness of the $\hat{\sigma}^z$-gate  when based on Raman
transitions with electric dipole coupling. 

\par The second type of
collective quantum gates, pioneered by S{\o}rensen and M{\o}lmer
\cite{Molmer-sorensen}, entangles the qubits by inducing collective
spin flips of both ions. It is based on a
$\hat{\sigma}^{\phi}$-dependent force, with $\hat{\sigma}^{\phi}$ being a linear combination of  the Pauli spin operators $\hat{\sigma}^{x}$ and $\hat{\sigma}^{y}$,
capable of operating on clock states \cite{Haljan}. Even
though this gate is formally equivalent to the $\hat{\sigma}^z$-gate in a
rotated basis though, it has  the disadvantage that it involves spin-flips,
 which make the gate more sensitive to
magnetic field fluctuations than the $\hat{\sigma}^z$-gate \cite{Leibfried1}. Spin flips also constitutes a drawback for quantum computation in decoherence-free subspaces protected against  collective dephasing, for which controlled-phase gates are naturally better suited \cite{Us}.  
Furthermore, the
required laser beam configuration gives rise to strong
sensitivity of the qubits' coherences on the optical phases of the
driving fields, constituting a serious limiting factor to the
fidelity of the gate \cite{Sackett} unless special spin echo
techniques are applied within the gate operation \cite{Haljan,
Lee}. In addition, for this type of gates spontaneous emission also sets a fundamental major limit to the fidelity.

\par In this paper, we
investigate how to overcome the $\hat{\sigma}^z$-gate's
spontaneous emission problem and its inefficiency with clock
states by using electric quadrupole transitions for mediating the
Raman coupling. The main idea is to induce a state-dependent force on the ions via  Raman lasers tuned close to a
narrow transition. We illustrate the idea with the $S_{1/2}\leftrightarrow D_{5/2}$ transition in $^{43}$Ca$^{+}$,  but it can also be applied to other ion species with a similar level structure \cite{otherspecies}. Two non-copropagating lasers are tuned such that the two qubit levels, $\ket{\!\uparrow}\equiv\ket{S_{1/2}(F=3,m_{F}=0)}$ and $\ket{\!\downarrow}\equiv\ket{S_{1/2}(F=4,m_{F}=0)}$, experience
opposite detunings from the metastable $D_{5/2}$ level (see Fig.~\ref{frequencies}  $b)$). Two ions placed inside the standing wave pattern created by such laser beams experience a
force. The direction and amplitude of the force on each ion depends jointly on the ions' internal state configuration, the distance between them, and the collective motional mode chosen. As in Ref.~\cite{Leibfried1}, a conditional dynamics is achieved by introducing a frequency difference of both beams close to the frequency a vibrational mode of the ion string. As a result, the state-dependent force becomes periodic and excites, for instance, the center-of-mass (CM) mode if and only if the two ions are in opposite internal states and separated by a semi-integer multiple of wavelengths. The periodicity induced by the detuning from the vibrational frequency ensures that the ion string eventually returns to its initial motional state. During this process, however, the ion string picks up a phase proportional to its excitiation, which allows to realize  a maximally entangling controlled-phase gate.

\par Distinguishing between the internal ionic levels only via
their respective detunings works for any pair of qubit levels with
a sufficiently large energy difference  regardless of their magnetic properties, in particular for
clock states. However, it requires that the mediator level  (the
$D_{5/2}$ level in our example) is long-lived, to achieve a
reasonably small spontaneous emission rate. Concentrating again on
$^{43}$Ca$^{+}$, the $D_{5/2}$ level has a lifetime of about 1~s, such that
the probability of spontaneous emission of a photon from both ions, and for a Raman
detuning of half the energy splitting of the groundstate hyperfine structure, is well below the
asymptotic threshold for fault-tolerant quantum computation found in \cite{Steane}.

{\it Spectral discrimination}. We now describe the emergence of the state dependent
force via spectral discrimination and the implementation of the
$\hat{\sigma}^z$-gate. For a more detailed explanation on general spin-dependent forces we refer the
reader to Ref.~\cite{Lee} and references therein.  We are interested in the implementation of
the interaction Hamiltonian
\begin{eqnarray}
\label{Hamilgeneral}
\hat{H}_{\text{int}_{i}}=
\sum_{m_{i}}\big(F_{0_{m_{i},i}}z_{\nu_{\text{cm}}}\hat{a}_{\text{cm}}^{\dagger}e^{i\delta t}+h.c.\big)\ket{m_{i}}\bra{m_{i}}|,\ \
\end{eqnarray}
with $m_{i}=\uparrow$ or $\downarrow$ being the pseudo-spin state of ion $i$,
for $i=1$ or $2$.  Hamiltonian
(\ref{Hamilgeneral}) describes a particle with internal states $\ket{m_{i}}$ trapped in a harmonic potencial of
frequency $\nu_{\text{cm}}$ and  driven by
a classical periodic force of amplitude $F_{0_{m_{i},i}}$ and
frequency $\delta\ll\nu_{\text{cm}}$. We consider the collective center-of-mass (CM) motional mode as the
 harmonic oscillator  because it exhibits lower levels of dephasing \cite{Roos}. Nevertheless, the treatment
can be straightforwardly applied to the stretch mode, too.  In Eq.~(\ref{Hamilgeneral}),
$z_{\nu_{\text{cm}}}\equiv\sqrt{\hbar/2M\nu_{\text{cm}}}$, where $M$ is the
single-ion mass, is the root mean square spread of the motional
ground state wave packet and $\hat{a}_{\text{cm}}^{\dagger}$ is the
creation operator of one phonon of the CM mode. Upon evolution with Hamiltonian~(\ref{Hamilgeneral}) the
oscillator is periodically driven through a circular trajectory in
phase space. For an evolution time $T=2n\pi\delta^{-1}$, with
$n\in\mathbb{N}$, it ends up back in its initial position, but
with each state $\ket{m_{i}}$ having acquired a geometrical phase $\Phi
(T=2n\pi\delta^{-1})\equiv
2n\pi|\frac{z_{\nu_{\text{cm}}}F_{0_{m_{i},i}}}{\delta\hbar}|^{2}$.
When the force is exerted simultaneously on both ions, the phase
eventually acquired by each composite state $\ket{m_{1},m_{2}}$ can be set to yield the
maximally-entangling $\hat{\sigma}^z$-gate by properly controlling the values of $F_{0_{m_{1},1}}$, $F_{0_{m_{2},2}}$ and $\delta$, as is described below.

\par
Fig.~\ref{frequencies} $b)$ sketches the relevant frequencies,
$\omega_{A}\equiv\omega_{\uparrow,e}+\Delta-(\nu_{\text{cm}}-\delta)$ and
$\omega_{B}\equiv\omega_{\uparrow,e}+\Delta$,
$\omega_{B}-\omega_{A}=\nu_{\text{cm}}-\delta$,  with
$\omega_{\uparrow,e}$ being the frequency difference between the
excited state $\ket{e}$ and $\ket{\!\uparrow}$, and $\omega_{0}$
between $\ket{\!\uparrow}$ and $\ket{\!\downarrow}$. We have
$\Delta\gg\nu_{\text{cm}}\gg\delta$, while  $\Delta$ and
$\omega_{0}$ are of the same order of magnitude.
Since the laser detuning $\Delta$ is
still much larger than the $D_{5/2}$ state's hyperfine
structure  we can treat the latter as an
effective single-level without internal
structure.
The individual interaction Hamiltonian in the Schr\"odinger picture and optical rotating-wave
approximation (RWA)
are  then given by
$\hat{H}'_{i}=\hbar\sum_{l,m_{i}}g_{l,i,m_i}\ket{e}\bra{m_i}|$
$e^{i[k_{l_{z}}\frac{z_{\nu_{\text{cm}}}}{\sqrt{2}}(\hat{a}_{\text{cm}}^{\dagger}+\hat{a}_{\text{cm}})
-\omega_{l}t-\varphi_{l}+k_{l_{z}}z_{0_{i}}]}+ H.c.$ \cite{comment0}. The Rabi frequency $g_{l,i,m_i}$, of laser  $l=A$ or $B$,
couples level $\ket{m_i}$ with $\ket{e}$ of ion $i$, $k_{l_{z}}$ is the
component of the $l$-th laser's wave vector along the trap axis $z$,
$\varphi_{l}$ its optical phase, and $z_{0_{i}}$ the equilibrium
position of ion $i$. 
If (I) $|g_{l,i,\uparrow}|\ll\Delta $ and $|g_{l,i,\downarrow}|\ll
|\Delta-\omega_{0}|$, and since $\Delta\approx|\Delta-\omega_{0}|$
is much larger than the excited state's linewidth, level
$\ket{e}$ can be adiabatically eliminated. The latter, together with
the RWA  eliminating terms rotating at frequencies $\Delta$ and
$\Delta-\omega_{0}$ yields the
following effective Hamiltonians in the interaction picture:
$\hat{H}'_{i}=
\hbar\sum_{m_{i}}\big[\chi_{m_{i},i}+(\theta_{m_{i},i}
e^{i[\eta_{\text{cm}}(\hat{a}_{\text{cm}}^{\dagger}e^{i\nu_{\text{cm}}t}+\hat{a}_{\text{cm}}e^{-i\nu_{\text{cm}}t})
-(\nu_{\text{cm}}-\delta) t-\phi_{i}]} 
+h.c.)\big]$ $\ket{m_{i}}\bra{m_{i}}|$.
Here $\eta_{\text{cm}}\equiv\Delta k_{z}\frac{z_{\nu_{\text{cm}}}}{\sqrt{2}}$, 
with $\Delta k_{z}\equiv k_{B_{z}}-k_{A_{z}}$, is the Lamb-Dicke
parameter and  $\phi_{i}\equiv\varphi_{B}-\varphi_{A}-\Delta
k_{z}z_{0_{i}}$.
$\chi_{\uparrow,i}\equiv-\frac{|g_{A,i,\uparrow}|^{2}+|g_{B,i,\uparrow}|^{2}}{\Delta}$
and
$\chi_{\downarrow,i}\equiv-\frac{|g_{A,i,\downarrow}|^{2}+|g_{B,i,\downarrow}|^{2}}{\Delta-\omega_{0}}$
are the time-averaged components of the AC Stark shift, and
$\theta_{\uparrow,i}\equiv-\frac{g_{B,i,\uparrow}g_{A,i,\uparrow}^{*}}{\Delta}$
and
$\theta_{\downarrow,i}\equiv-\frac{g_{B,i,\downarrow}g_{A,i,\downarrow}^{*}}{\Delta-\omega_{0}}$
its time-dependent components.
\par
The last Hamiltonian still contains fast oscillations at
frequency $\nu_{\text{cm}}$. In (II) the resolved
side-band limit
$|\theta_{m_{i},i}|\ll\nu_{\text{cm}}$  only
the stationary (non-oscillating) terms give a significant
contribution to the effective Hamiltonian, whereas the other can
be neglected under the RWA. Furthermore,   (III) in the
Lamb-Dicke limit (LDL) $\eta_{\text{cm}}^{2}(n_{\text{cm}}+1/2)\ll 1$, where
$n_{\text{cm}}$ is the mean population of the CM mode, only the terms
linear in $\eta_{\text{cm}}$ contribute. Then the effective
Hamiltonians are finally given by:
\begin{eqnarray}
\nonumber
\hat{H}_{\text{int}_i}=\hbar\sum_{m_{i}}\big[\chi_{m_{i},i}\\
+\big(i\theta_{m_{i},i}\eta_{\text{cm}}e^{-i\phi_{i}}\hat{a}_{\text{cm}}^{\dagger}e^{i\delta t}+h.c.\big)\big]\ket{m_{i}}\bra{m_{i}}| .
\label{hamintinLDL}
\end{eqnarray}
\par
Let us momentarily disregard the time-averaged Stark shift from Hamiltonian~(\ref{hamintinLDL}) and consider only its
time-dependent component that contains the vibrational operators
generating the interaction between both qubits. Under this
assumption it  suffices to define $F_{0_{m_{i},i}}z_{\nu_{\text{cm}}}\equiv
i\hbar\theta_{m_{i},i}\eta_{\text{cm}}e^{-i\phi_{i}}$ to identify
Hamiltonian~(\ref{hamintinLDL}) with (\ref{Hamilgeneral}), realizing
thus the desired $\hat{\sigma}^z$-dependent force on both ions. For
simplicity, we take $g_{l,1,m}=g_{l,2,m}\equiv
g_{l,m}\Rightarrow\theta_{m,1}=\theta_{m,2}\equiv\theta_{m}\Rightarrow|F_{0_{m,1}}|=|F_{0_{m,2}}|\equiv
|F_{0_{m}}|$, as no individual laser addressing has been assumed. The extension to non-equal coupling though is straightforward. The phase
difference between the forces applied to both ions is then
determined by the ion spacing at equilibrium:
$\phi_{1}-\phi_{2}=\Delta k_{z}\Delta z_{0}$, with $\Delta
z_{0}\equiv z_{0_{1}}-z_{0_{2}}$. Here it is convenient to set the
spacing equal to a semi-integer multiple of the optical wavelength:
$\Delta k_{z}\Delta z_{0}=(2n+1)\pi$, so that both ions experience opposite forces. In this way the total
force on the CM mode $F_{0_{m_{1},m_{2}}}z_{\nu_{\text{cm}}}\equiv
(F_{0_{m_{1}}}+F_{0_{m_{2}}})z_{\nu_{\text{cm}}}=\hbar
i(\theta_{m_{1}}-\theta_{m_{2}})\eta_{\text{cm}}e^{-i\phi_{1}}$ is non-vanishing
only for anti-aligned spins.
\par
After one phase-space round trip of duration $T=2\pi\delta^{-1}$ (the shortest
gate-time possible) the composite states evolve as:
$\ket{m_{1},m_{2}}\rightarrow\ket{m_{1},m_{2}}$, for
$m_{1}=m_{2}$; and  $\ket{m_{1},m_{2}}\rightarrow
e^{i\Phi}\ket{m_{1},m_{2}}$, for $m_{1}\neq m_{2}$, with
$\Phi=2\pi|(\theta_{\uparrow}-\theta_{\downarrow})\eta_{\text{cm}}\delta^{-1}|^{2}$.
Maximal entanglement is achieved for $\Phi=\frac{\pi}{2}$; and the
bigger the difference between $\theta_{\uparrow}$ and
$\theta_{\downarrow}$, the faster the gate. Imposing thus $\Phi=\frac{\pi}{2}$ and using
the definition of $\theta_{m}$, the
general discrimination condition for maximal entanglement is obtained:
\begin{equation}
\label{SelectivityCondition}
\bigg|\frac{g_{B,\uparrow}g_{A,\uparrow}^{*}}{\Delta}-\frac{g_{B,\downarrow}g_{A,\downarrow}^{*}}{\Delta-\omega_{0}}\bigg|=\bigg|\frac{\delta}{2\eta_{\text{cm}}}\bigg|\, .
\end{equation}
Since we are most interested in working with qubit states with the same magnetic
properties no discrimination via polarization can
take place, {\it i. e.}, $g_{l,\uparrow}=g_{l,\downarrow}\equiv
g_{l}$, for $l=A$ or $B$. Therefore, the only way to fulfill
condition~(\ref{SelectivityCondition}), which then reads
$\big|g_{B}g_{A}^{*}\big[\Delta^{-1}-(\Delta-\omega_{0})^{-1}\big]\big|=\big|\frac{\delta}{2\eta_{\text{cm}}}\big|$,
is through spectral discrimination. Maximal discrimination occurs
at $\Delta=\frac{\omega_{0}}{2}$
, when
$\theta_{\uparrow}=-\theta_{\downarrow}$.  As
shown in figure~\ref{frequencies} $b)$, for this detuning,
$\ket{\uparrow}$ couples to the excited state $\ket{e}$ through
blue-detuned light and $\ket{\downarrow}$ through red-detuned
light of the same intensity. So both states' effective couplings
with the motional sideband (and therefore also the forces each
spin state experiences) are exactly opposite. Thus the optimal
form of condition~(\ref{SelectivityCondition}) is
\begin{equation}
\label{SpectralselectivityCondition}
\big|g_{B}g_{A}^{*}\big|=\bigg|\frac{\delta\omega_{0}}{8\eta_{\text{cm}}}\bigg|\, .
\end{equation}

\par
For $^{43}$Ca$^{+}$, the ground state hyperfine splitting is
$\omega_{0}=2\pi\times 3.226$ GHz. The resulting Raman detuning is  20 times larger than
the hyperfine splitting of the $D_{5/2}$ manifold, so that the
single-level approximation is comfortably satisfied. It is also much
larger than the linewidth $\gamma_{D}=2\pi\times$0.18~Hz. For the
typical experimental parameters: $\nu_{\text{cm}}=2\pi\times 1.2$~MHz,
$\eta_{\text{cm}}=0.1$ and $\delta=2\pi\times 1$~kHz, and taking
$|g_{A}|=|g_{B}|\equiv|g|$, the value $g=2\pi\times 2$ MHz is
obtained from Eq.~(\ref{SpectralselectivityCondition}) for the optical quadrupole couplings. All the other
approximations made in the derivation of
Hamiltonian~(\ref{hamintinLDL}) are in turn well satisfied. It is:
(I) $|g|=2\pi\times 2$~MHz $\ll 2\pi\times
1.613$~GHz$=|\Delta-\omega_{0}|=\Delta$;  (II)
$\frac{|\theta_{m}|}{\nu_{\text{cm}}}=0.002\ll 1$; and (III)
$\eta_{\text{cm}}^{2}(n_{\text{cm}}+1/2)=0.01(n_{\text{cm}}+1/2)$, which is much smaller
than 1 for the mode populations of interest. On the other hand, the
total duration of the gate (limited by the available amount of laser power) is $T=$1 ms, which is short as compared
to the long coherence times expected for clock states \cite{Langer}. To speed up the gate time to 100 $\mu$s for example, a total laser power of about 1 W, when focussing the laser beams to waist sizes of 6 $\mu$m, would be required.

\par
Not all types of spontaneous emission affect the gate's performance equally  \cite{Ozeri2}. The dominant source of gate error comes from inelastic Raman scattering that affects the ion-qubit coherences directly, whereas elastic Rayleigh scattering can only affect the gate's fidelity indirectly through the ions' recoils. In what follows, we calculate the total probability of spontaneous photon scattering and simply take it as an overestimated figure of merit for the gate error.
The mean probability of off-resonant direct excitation of the
excited state $\ket{D_{5/2}}$ under the action of the two
$\frac{\omega_{0}}{2}$-detuned lasers is approximately
$p_{\text{off}}=8\frac{|g|^2}{\omega_{0}^2}\equiv 2\times 10^{-6}$. The
total probability of spontaneous emission from both ions during the
gate-time $T=2\pi\delta^{-1}$ is then: $p_{\text{T}}\equiv 2\times
p_{\text{off}}\times\gamma_{D}\times 2\pi\delta^{-1}\equiv  2\times
8\frac{|g|^2}{\omega_{0}^2}\times\gamma_{D}\times\frac{2\pi\omega_{0}}{8|g|^2
\eta_{\text{cm}}}\equiv\frac{4\pi}{\eta_{\text{cm}}}\frac{\gamma_{D}}{\omega_{0}}\equiv
6.3\times 10^{-9}$, where
condition~(\ref{SpectralselectivityCondition}) has been used. This
value is more than four orders of magnitude below the asymptotic
threshold value set in \cite{Steane} for fault-tolerant quantum
computation. When it comes to choosing a specific ion as a quantum information carrier in the case of gate operations mediated by a short-lived $P$ level, it seems favourable to avoid ion species with low-lying $D$ levels  \cite{Ozeri2}. The gate scheme proposed here not only works for any ion species with a low-lying narrow transition, but also it is exactly the use of the low-lying narrow transition as the mediator what enables an error threshold four orders of magnitude smaller than for gates mediated by short-lived $P$ levels.

\par
{\it Single-ion phase compensation}. Let us now go back to the time-averaged Stark shifts disregarded from Hamiltonian~(\ref{hamintinLDL}). In principle, the single-ion phases they induce can be compensated at the end of the evolution by single-ion gates. As for the robustness of an implementation however, they are very important. Since $|\chi_{m_i}|$ is about twenty times as big as $|\theta_{m_i}\eta_{\text{cm}}|$,  the acquired single-ion phases are much larger than the desired conditional $\pi/2$ phase, which  makes the gate's performance extremely sensitive to laser-intensity fluctuations. One way to overcome this
is to add additional frequencies onto the gate-laser so as to compensate the time-averaged components of the Stark shift without compensating its time-varying components \cite{Haeffner03}. Alternatively, a spin-echo-technique can be applied. In this approach, the gate is divided  into two phase-space round trips, in each of which half the desired conditional phase is acquired, with a pulse interchanging the qubit populations in the middle. This increases the gate time by a factor of $\sqrt{2}$, but for laser power fluctuations slower than the gate time both ions pick up the same final single-ion phase. These two techniques can also be combined.

\par {\it Discussion}. Since the scheme is based on the use of weak transitions, the large amount of
laser power it requires to achieve short gate times is somewhat challenging.  In a proof-of-principle experiment this technical drawback can be circumvented using alternative qubit encodings that decrease the required laser power. For instance, one can map the qubit onto clock states of the hyperfine splitting of the $D_{5/2}$ manifold. Hereby the qubit transition frequency is reduced by
almost a factor of 500, so that the required laser couplings are reduced by more than one order of magnitude. A ground state sublevel  serves as the Raman mediator and coupling to the other Zeeman sublevels of the ground state manifold can be avoided by properly choosing the laser
polarizations.  Since the $D_{5/2}$ level is now occupied by both qubit states throughout the gate, the probability of spontaneous emission is directly given by $p_{\text{T}}\equiv 2\times\gamma_{D}\times T$. For instance, for an estimated gate time of $T=100$~$\mu$s ($g\approx2\pi\times 285$~kHz) including mapping of the qubits, the spontaneous scattering per gate-run would still be as low as $p_{\text{T}}= 2\times 10^{-4}$. 
Alternatively, another possible encoding is provided by one clock state from $S_{1/2}$ and another from $D_{5/2}$, where values of $g\approx 2\pi\times 210$~kHz and $p_{\text{T}}\approx 10^{-4}$ are expected \cite{Kihwan}. 

\par {\it Summary}. We propose  a new implementation of the $\hat{\sigma}^z$-gate on hyperfine ground-state qubits that overcomes the two main problems of previous implementations: spontaneous emission and inefficiency with clock states. The gate is driven by forces  exerted by two beams in a Raman configuration tuned close to a narrow transition, and state
dependency is achieved by a small frequency detuning instead of polarizations. We discuss the idea  for  $^{43}$Ca$^{+}$ but it can also be applied to other ion species with a similar level structure. We
find a total probability of spontaneous emission per gate-run of less than 10$^{-8}$, thus eliminating
the main source of gate infidelity of previous implementation with magnetic-field-sensitive states. In order to achieve short gate times, the scheme requires a large amount of laser power. We show two examples of how to circumvent this by using alternative qubit encodings. With increasingly-powerful laser sources though, our $\hat{\sigma}^z$-gate can set a standard for robust high-fidelity entanglement creation in the future large-scale ion-trap quantum computer, where the qubit encoding will be that of hyperfine ground-state sublevels. 
With this scheme it seems possible to reach the fault tolerant threshold. 

\begin{acknowledgments}
\par This work was financially supported by CAPES/DAAD, FAPERJ, the Brazilian
Millennium Institute for Quantum Information, the Austrian Science
Fund (FWF), the European Commission (CONQUEST, SCALA
networks) and the Institut f\"ur Quanteninformation GmbH.
\end{acknowledgments}


\begin{thebibliography}{99}
\bibitem{comment0}
In this Hamiltonian we have already disregarded all operators associated to the stretch motional mode, as for our frequency configuration they will give no stationary contribution and can be neglected in the resolved side-band limit with the RWA.
\end{thebibliography}
\end{document}